\documentclass{jaa}
\usepackage[numbers]{natbib}
\usepackage{graphicx}
\usepackage{amsmath, amssymb, amsfonts}
\usepackage{bm} 
\usepackage{cuted}
\usepackage{multirow}
\usepackage{array}

\begin{document}\sloppy

\title{An interesting equation leads to the MOND but does not rule out the existence of dark matter}



\author{Jinwen Hu\textsuperscript{1,*}}
\affilOne{\textsuperscript{1}Department of physics and technology, Wuhan university, Wuhan 430060, China}


\twocolumn[{

\maketitle

\corres{200731890025@whu.edu.cn}

\msinfo{10 November 2026}{15 May 2026}

\begin{abstract}
Inspired by the idea in Ref. [16], which introduced a viscosity coefficient into the $\Lambda \mathrm{CDM}$ model to describe the expansion of universe, driven by curiosity, we also attempt to introduce such a positive viscosity coefficient into the rotational motion equation describing the disk galaxies, despite the physical spatial scale of the two objects of study are different, and then study what will happen. Coincidentally and surprisingly, we found that this equation can derive all the formulas assumed in MOND, especially the concrete interpolation function between the centripetal acceleration and the Newtonian acceleration, which however is empirical in MOND. But at the same time, something different from MOND was also obtained, that is, the critical acceleration, $a_{0}$ in MOND, does not need to be a constant, but increases with the mass of the galaxy increases, and with the action of viscosity coefficient, the rotational galaxies will gradually expand radially over time at an extreme small expansion rate, just like the expansion of universe. However, unlike MOND, the model in this paper cannot rule out the existence of dark matter assumed in $\Lambda \mathrm{CDM}$ (in fact, we tend to consider the idea of this paper to be a further optimization of $\Lambda \mathrm{CDM}$ rather than an alternative to $\Lambda \mathrm{CDM}$ ). Instead, the mass of dark matter can be used to help to adjust the value of $A_{0}$ (here it just to distinguish from $a_{0}$ in MOND, and $A_{0}$ and $a_{0}$ in the equation have the same meaning), thereby helping to better fit the radial acceleration relation (RAR) curve of galaxies. However, unlike $\Lambda \mathrm{CDM}$, even if dark matter exists, it does not need to be carefully adjusted to meet the asymptotically flat rotational velocity curve of disk galaxies, which is considered to be unnatural by Milgrom and leading to the proposal of MOND. And the rotational curve of disk galaxies with this characteristic can be also achieved naturally under the viscous dynamics of the galaxy itself.
\end{abstract}


}]


\doinum{12.3456/s78910-011-012-3}

\artcitid{}
\volnum{}
\year{}
\pgrange{1--}
\setcounter{page}{1}
\lp{1}



\section{Introduction}
Dark matter and dark energy are the two major puzzles in cosmology today[1]. Dark energy is proposed to explain why universe is accelerating, and dark matter is proposed to solve the abnormal rotation curve of disk galaxies. In the standard cosmological model, $\Lambda$ Cold Dark Matter ( $\Lambda \mathrm{CDM}$ ), the component of dark energy is about $70 \%$, the component of dark matter is about $25 \%$, and the other component we can see is barely about 5\%[2]. Although the standard cosmological model is relatively successful and accepted by most physicists, it still has some unresolved problems, such as the fine-tuning problem and the origin problem of dark energy[3]. And for dark matter, people are still not sure what the component it is[4], since none of the particles described or predicted by the Standard particle Model meet the properties of dark matter. And more importantly, the dark matter have not been detected by experiments so far. It is in this complex context that various alternative models on Dark Energy and Dark Matter have been proposed. And among these alternative models to Dark Matter, the well-known and famous model maybe the MOdified Newtonian Dynamics (MOND)[5], which is first proposed by Milgrom in 1983. Milgrom postulates that at small accelerations the usual Newtonian dynamics break down and that the law of gravity needs to be modified. To date, MOND have been widely studied and it has been shown to successfully explain a large number of astrophysical observations, such as kinematics of galaxies[6,7], dynamics of wide binary stars[8,9], radial acceleration relation of galaxies[10,11], orbital velocity of interacting galaxy pairs[12], early-universe galaxy and cluster formation[13], and galaxy-scale gravitational lensing[14,15]. However, for alternative models to Dark Energy, they are still difficulty to distinguish, and so far no alternative model has stood out, but we notice an interesting model proposed in Ref. [16], which argues that the cosmic fluid is not perfect but rather viscous and dissipative, thus introducing a viscous pressure proportional to the Hubble constant into the standard cosmological model. Although the origin of the viscous pressure is still debated, it shows that the bulk viscosity model is significantly better than the $\Lambda \mathrm{CDM}$ model in fitting the combined $\mathrm{SNe} \mathrm{Ia}+\mathrm{CMB}+\mathrm{BAO}+\mathrm{H}(\mathrm{z})$ data.

In this paper, inspired by the idea in Ref. [16] that introducing a dissipation term into the $\Lambda \mathrm{CDM}$ model to describe the expansion of universe, driven by curiosity, we also try to introduce such a dissipation term into the rotational motion equation describing the disk galaxies, and then study its effect on the rotational motion of galaxies. Surprisingly, we found a high degree of coincidence between the derived motion properties and those predicted by MOND. However, as we have known that, MOND is only an empirical model, not a fundamental theory[14]. This seems to give us a window to glimpse the nature of MOND, or in other words, even if MOND is not a correct model, but as MOND can explain so many astronomical phenomena, we can also glimpse the secrets hidden behind MOND.

\section{Newtonian dynamics with dissipation term}
In Ref. [16], the author introduced a dissipation term into the $\Lambda \mathrm{CDM}$ model to describe the expansion of universe, making the expansion and evolution equation describing the universe becomes
\begin{equation}\label{eq01}
\ddot{a}=-\frac{4 \pi G \rho_{m}}{3} a+\frac{\Lambda}{3} a+12 \pi G \zeta \dot{a} 
\end{equation}
Where $a$ denotes the scale factor of universe, $\rho_{m}$ is the total matter density, $G$ is the gravitational constant, $\Lambda$ denotes the cosmological constant, $\zeta$ denotes the introduced viscosity coefficient, which is positive.

In Ref. [16], it shows that Eq. (\ref{eq01}) is significantly better than the $\Lambda \mathrm{CDM}$ model in fitting the astrophysical observation data. Inspired by this, we also try to introduce such a dissipation term into the rotational motion equation describing the disk galaxies and study what will happen, even if the physical basis of this dissipation term is unclear at present, and even if the physical spatial scales of the two objects of study are different.

Similar to Eq. (\ref{eq01}), we also assume that there is a dissipation term in the rotational motion equation describing the disk galaxy that is proportional to the velocity, and that the corresponding viscosity coefficient is also positive just as which in Eq. (\ref{eq01}). Then, without considering relativistic effects, we can easily obtain the force equilibrium equations in the radial and circumferential directions of rotational motion
\begin{equation}\label{eq02}
 \left\{\begin{array}{l}
\ddot{r}+\frac{G M}{r^{2}}=\dot{\theta}^{2} r+\lambda \dot{r}  \\
\frac{d}{d t}(\dot{\theta} r \cdot r)=\lambda \dot{\theta} r \cdot r
\end{array}\right.
\end{equation}
Where $M$ denotes the center mass of the disk galaxy, $(r, \theta)$ denotes the coordinates of motion and here for simplicity we only consider the motion in a two-dimensional plane, $\lambda$ is the assumed viscosity coefficient. And the second formula of Eq. (\ref{eq02}) represents that the derivative of angular momentum corresponding to the rotational motion with respect to time is equal to the rotational torque.

One may notice that, in Eq. (\ref{eq02}) it also implies an assumption that the dissipation force is related to the mass $m$ of the object in rotating motion, because in its first formula both sides of the equation cancel out the mass $m$, which is also consistent with the idea in Ref. [16]. In fact, Eq. (\ref{eq02}) is equivalent to introducing a dissipation function $\psi$ into the Newtonian dynamics
\begin{equation}\label{eq03}
\psi=-\frac{1}{2} \lambda m\left[\dot{r}^{2}+(r \dot{\theta})^{2}\right]
\end{equation}
 
Now anyway, we can use Eq. (\ref{eq02}) to study the rotational motion of disk galaxies. First from the second formula of Eq. (\ref{eq02}) we can obtain
\begin{equation}\label{eq04}
 \dot{\theta} r^{2}=\dot{\theta_{0}} r_{0}^{2} e^{\lambda t}
\end{equation}
where $\left(\theta_{0}, r_{0}\right)$ is the initial condition. Then substituting Eq. (\ref{eq04}) and $G M=v_{0}{ }^{2} r_{0}$ into the first formula of Eq. (\ref{eq02})
\begin{equation}\label{eq05}
\ddot{R}=\lambda \dot{R}+\frac{v_{0}^{2}}{r_{0}^{2}} e^{2 \lambda t} \frac{1}{R^{3}}-\frac{v_{0}^{2}}{r_{0}^{2}} \frac{1}{R^{2}}
\end{equation}
where $R=r / r_{0}$ is a dimensionless quantity, and $v_{0}=\dot{\theta_{0}} r_{0}$. Here it should be noted that, in the derivation of Eq. (\ref{eq05}), we used the equation $G M=v_{0}{ }^{2} r_{0}$, which assumes that the object in rotating motion is in a strong gravitational field at initial, thus leading to Eq. (\ref{eq02}) returns to the usual Newtonian dynamics.

\begin{table*}[ht]
\centering
\caption{The numerical analysis of Eq. (\ref{eq05})}\label{tab01}
\begin{tabular}{|>{\centering\arraybackslash}p{2cm}|
                >{\centering\arraybackslash}p{3cm}|
                >{\centering\arraybackslash}p{4cm}|
                >{\centering\arraybackslash}m{3cm}|
                >{\centering\arraybackslash}m{3cm}|}
\hline
Serial Number & The assumed value of $\lambda$ & The assumed initial condition ($v_{0}, r_{0}$) & \multirow{2}{*} {The obtained $A_{0}$ }& \multirow{2}{*} {The obtained $\mu^{\prime}$} \\
\hline
1 & $\lambda=1 \times 10^{-7}$ & $v_{0}=1$, $r_{0}=1$ & $1.0038 \times 10^{-10}$ & \\
2 & $\lambda=1 \times 10^{-7}$ & $v_{0}=10$, $r_{0}=1$ & $4.7323 \times 10^{-10}$ & \\
3 & $\lambda=1 \times 10^{-7}$ & $v_{0}=100$, $r_{0}=1$ & $2.1986 \times 10^{-9}$ & \\
4 & $\lambda=1 \times 10^{-7}$ & $v_{0}=1000$, $r_{0}=1$ & $1.02257 \times 10^{-8}$ & \\
5 & $\lambda=1 \times 10^{-7}$ & $v_{0}=0.1$, $r_{0}=1$ & $2.0005 \times 10^{-11}$ & \\
6 & $\lambda=1 \times 10^{-7}$ & $v_{0}=10$, $r_{0}=0.01$ & $1.0021 \times 10^{-10}$ & \\
7 & $\lambda=1 \times 10^{-7}$ & $v_{0}=50$, $r_{0}=0.0004$ & $1.0028 \times 10^{-10}$ & \\
8 & $\lambda=1 \times 10^{-7}$ & $v_{0}=100$, $r_{0}=0.0001$ & $1.0025 \times 10^{-10}$ & \\
9 & $\lambda=5 \times 10^{-7}$ & $v_{0}=1$, $r_{0}=1$ & $8.6415 \times 10^{-10}$ & \\
10 & $\lambda=5 \times 10^{-7}$ & $v_{0}=10$, $r_{0}=0.01$ & $8.6519 \times 10^{-10}$ & \\
11 & $\lambda=5 \times 10^{-7}$ & $v_{0}=50$, $r_{0}=0.0004$ & $8.6945 \times 10^{-10}$ & See Figure \ref{fig01} \\
12 & $\lambda=5 \times 10^{-7}$ & $v_{0}=100$, $r_{0}=0.0001$ & $8.7017 \times 10^{-10}$ & \\
13 & $\lambda=1 \times 10^{-6}$ & $v_{0}=1$, $r_{0}=1$ & $2.1844 \times 10^{-9}$ & \\
14 & $\lambda=1 \times 10^{-6}$ & $v_{0}=10$, $r_{0}=0.01$ & $2.1879 \times 10^{-9}$ & \\
15 & $\lambda=1 \times 10^{-6}$ & $v_{0}=50$, $r_{0}=0.0004$ & $2.1902 \times 10^{-9}$ & \\
16 & $\lambda=1 \times 10^{-6}$ & $v_{0}=100$, $r_{0}=0.0001$ & $2.1931 \times 10^{-9}$ & \\
17 & $\lambda=1 \times 10^{-6}$ & $v_{0}=100$, $r_{0}=1$ & $4.7171 \times 10^{-8}$ & \\
18 & $\lambda=2 \times 10^{-6}$ & $v_{0}=100$, $r_{0}=1$ & $1.1933 \times 10^{-7}$ & \\
19 & $\lambda=5 \times 10^{-6}$ & $v_{0}=10$, $r_{0}=0.01$ & $1.8766 \times 10^{-8}$ & \\
20 & $\lambda=5 \times 10^{-6}$ & $v_{0}=100$, $r_{0}=0.0001$ & $1.8777 \times 10^{-8}$ & \\
21 & $\lambda=5 \times 10^{-6}$ & $v_{0}=1000$, $r_{0}=1 \times 10^{-6}$ & $1.8776 \times 10^{-8}$ & \\
\hline
\end{tabular}
\end{table*}

The analytical solution of Eq. (\ref{eq05}) is difficult to obtain, but we can obtain its numerical solution. Since there is currently no clear physical basis to help us to determine the value of $\lambda$ (from the success of Newtonian dynamics, we know that $\lambda$ must be an extreme small value), let us discuss the motion properties predicted by Eq. (\ref{eq02}) by arbitrarily setting the value of $\lambda$ and the initial condition of the equation. And similar to MOND, here, the motion properties we are interested in is also the relationship between the gravitational acceleration and the centripetal acceleration subjected by the object in rotating motion. And it is well known that in MOND, the relationship between the gravitational acceleration and the centripetal acceleration was assumed as
\begin{equation}\label{eq06}
g_{N}=a \mu\left(a / a_{0}\right)
\end{equation}
where $g_{N}$ denotes the gravitational acceleration, $a$ denotes the centripetal acceleration, $\mu$ is an empirical or undetermined interpolation function, and when $a / a_{0} \gg 1, \mu \rightarrow 1$, when $a / a_{0} \ll 1, \mu \rightarrow a / a_{0}$, and $a_{0} \approx 1.2 \times 10^{-10} \mathrm{~m} / \mathrm{s}^{2}$ is a constant.

As we know, in MOND, when the gravitational field is very weak, i.e., $a / a_{0} \ll 1$, there is $v_{u}{ }^{4}=G M a_{0}$, where $v_{u}$ denotes the circumferential velocity of the object in rotating motion, which can explain the asymptotically flat rotational velocity curve of disk galaxies. Coincidentally, we also find that based on Eq. (\ref{eq05}), no matter what the initial conditions are, as long as time evolves long enough, although the radial distance of the object in rotating motion is increasing over time, the circumferential velocity remains unchanged (here we assume the "unchanged" velocity is $v_{\text {ultimate }}$, and in the subsequent analysis, we will find that from Eq. (\ref{eq05}) it can derive all the formulas in MOND, which in turn proves this conclusion). Thus, similar to MOND, here we define a physical quantity $A_{0}$, and it satisfies
\begin{equation}\label{eq07}
v_{\text {ultimate }}^{4}=G M A_{0}
\end{equation}
Where $A_{0}$, which just to distinguish from $a_{0}$ in MOND, has the same meaning as $a_{0}$.

So next we will discuss Eq. (\ref{eq02}) or Eq. (\ref{eq05}) in terms of the formula form in MOND, that is
\begin{equation}\label{eq08}
g_{N}=a \mu^{\prime}\left(a / A_{0}\right)
\end{equation}

Now we will conduct numerical analysis on Eq. (\ref{eq05}), and Table \ref{tab01} shows the value of $A_{0}$ and $\mu^{\prime}$ corresponding to different $\lambda$ and initial conditions. Here we did not set units in the data in Table \ref{tab01}, because it is not necessary. Whether there are units in the numerical analysis does not affect properties or performance of Eq. (\ref{eq05}).

\begin{figure}[!ht]
\includegraphics[width=\columnwidth]{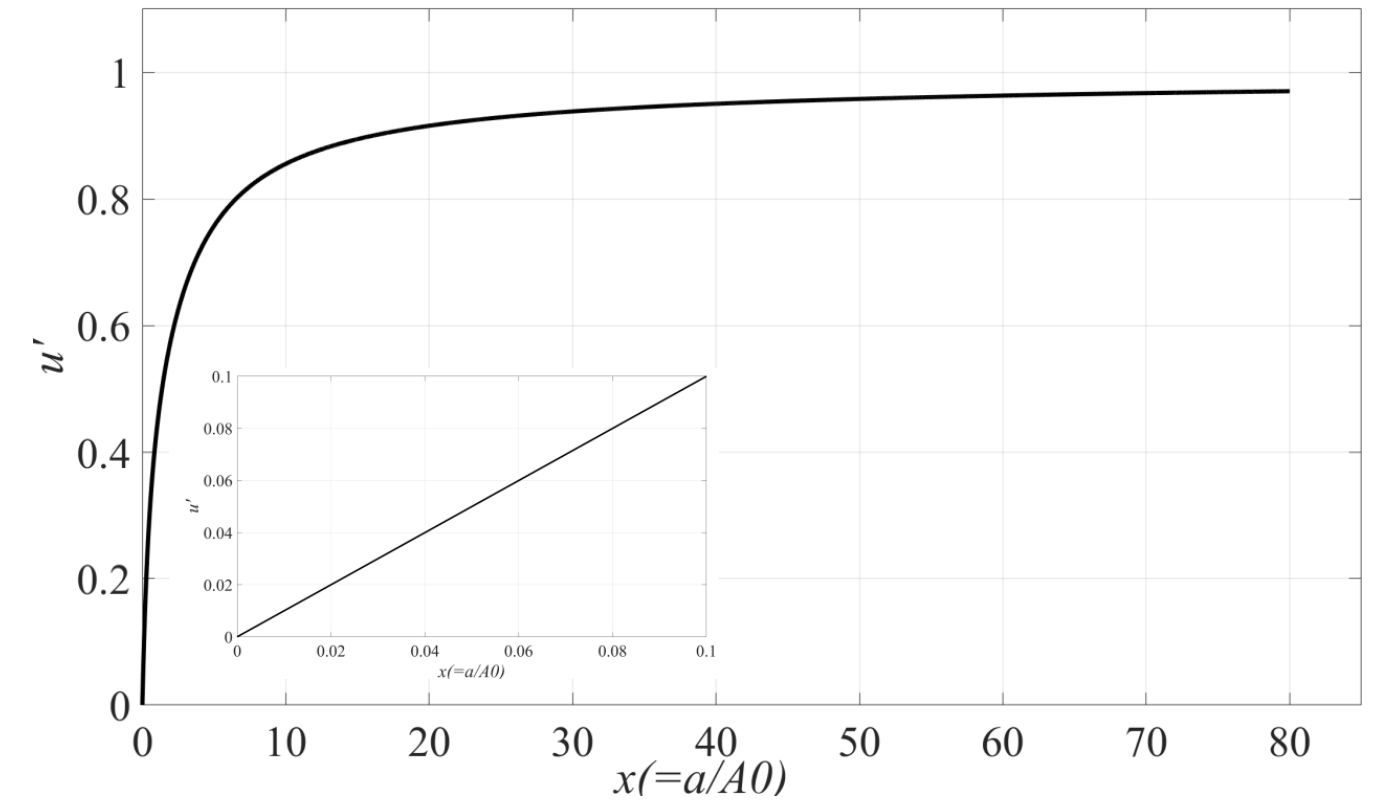}
\caption{The obtained $\boldsymbol{\mu}^{\prime}$ corresponding the values in Table \ref{tab01}.}\label{fig01}
\end{figure}

In Table \ref{tab01}, the number group (1,6,7,8), (9, 10,11,12), (13,14,15,16) and (19,20,21) corresponds to the same mass $M$, that is $G M=v_{0}^{2} r_{0}$. And from these number group we can see that $A_{0}$ is only dependent on $M$ and $\lambda$, not on the initial conditions $\left(v_{0}, r_{0}\right)$, as long as the initial centripetal acceleration corresponding to the initial conditions is strong enough, which is an important step for Eq. (\ref{eq02}) yields the MOND. One might wonder why does it here say that as long as the initial centripetal acceleration is strong enough, it's because if the initial centripetal acceleration $a^{\prime}{ }_{0}$ is too small and $\lambda$ is relatively large, $v_{\text {ultimate }}$ will be close to $v_{0}$, and therefore $A_{0} \approx a_{0}{ }^{\prime}$ according to Eq. (\ref{eq07}), which means that $A_{0}$ is dependent on $M$ or the initial conditions $\left(v_{0}, r_{0}\right)$, almost not on $\lambda$. In other words, if a large number of objects rotating around the center of the disk galaxy are initially in the strong gravitational potential region of the galaxy, and we assume that for this galaxy the value of $\lambda$ is a constant, then, regardless of which orbits these objects are initially distributed in (i.e., their orbits are different), they will all eventually reach the same rotational circumferential velocity under the action of dissipation function (i.e. Eq. (\ref{eq03})), and the fourth power of this ultimate rotational circumferential velocity (i.e., $v_{\text {ultimate }}$ ) is proportional to the central mass of the galaxy (i.e. Eq. (\ref{eq07}), which also known as the famous Tully-Fisher relation [17-19]).

In Table \ref{tab01}, from the number group (1,2,3,4,5) we can see that if we assume the value of $\lambda$ is a constant independent of the central mass of the disk galaxy, then, the larger the mass of a galaxy, the higher the value of $A_{0}$, which seems to echo to a model called EMOND (i.e., extend MOND, which assumes that $a_{0}$ in MOND is increased within the galaxy cluster, since that the MOND paradigm has achieved success on the galaxy scale, but it is unable to explain the galaxy clusters) [20], and we intend to further discuss this conclusion in the latter section.

In Table \ref{tab01}, from the number group (1,9,13) and (6,14,19) we can see that for the same galaxy, the higher the assumed value of $\lambda$, the higher the value of $A_{0}$.

At last, the most important conclusion from Table \ref{tab01} and Figure \ref{fig01} is that, as long as the initial centripetal acceleration corresponding to the initial conditions is strong enough, whatever the assumed value of $\lambda$ is, whichever the initial orbit of the object rotating around the center of the disk galaxy is, or whatever the center mass of the disk galaxy is, $\mu^{\prime}$, the interpolation function, are or are almost the same. And from Figure \ref{fig01} we can see that, when $x \ll 1, \mu^{\prime} \rightarrow x$, and when $x \gg 1, \mu^{\prime} \rightarrow 1$, which is exactly what is assumed in the MOND's empirical formula. Further, we find that the curve in Figure \ref{fig01} can be fitted as the following function

\begin{equation}\label{eq09}
\begin{aligned}
 \mu^{\prime}(x)=&0.1455 \frac{\sqrt{1+4 x}-1}{\sqrt{1+4 x}+1}-0.09382 \frac{x}{\sqrt{1+x^{2}}}\\
 &-0.2126\left(1-e^{-x}\right)+1.16092 \frac{x}{1+x}    
\end{aligned}
\end{equation}

Unfortunately, we here cannot give a clear mathematical definition to the above statement "as long as the initial centripetal acceleration is strong enough", mainly because for Eq. (\ref{eq05}) we cannot obtain its analytical solution, but its numerical solution.

In above, we only introduced a viscosity coefficient $\lambda$ into the Newtonian dynamics and then we have almost completely derived all the formulas assumed in MOND, especially the concrete interpolation function. However, we also obtained something different from MOND, that is, $A_{0}$ does not need to be a constant, and the disk galaxies are also gradually expanding over time at an extreme small expansion rate, just like universe is expanding.

Although the above conclusions are drawn based on the data in Table \ref{tab01}, the amount of data in Table \ref{tab01} is far from enough. In fact, we privately expanded the amount of data in Table \ref{tab01}, and found that the obtained conclusions remain the same, so we did not list these lengthy data. In addition, it should be stressed again that the absence of units for the data in Table \ref{tab01} does not affect the above obtained conclusions from Eq. (\ref{eq05}), or in other words, in Table \ref{tab01}, whether the unit of $\lambda$ is $1 / s$ or $1 / h$ or 1/year (the possible value of $\lambda$ will be discussed in Sect. 5), and whether the unit of $v_{0}$ is $m / s$ or $k m / s$, and whether the unit of $r_{0}$ is $m$ or parsec, Eq. (\ref{eq09}) remains the same, and the relationship between $A_{0}$ and $M$ remains the same.

\section{Discussion on $\boldsymbol{a}_{\mathbf{0}}$ from observations}
MOND, as a competitive model against $\Lambda \mathrm{CDM}$, was initially proposed with the aim of explaining the abnormal rotational velocity curves of galaxies. For MOND, the usual method is to fix the parameter $a_{0}$ and take $M / L$ (i.e., Mass-to-light ratio) as a free parameter, then based on the observational data, an optimal fitting parameter $M / L$ can be obtained. If we check the obtained parameter $M / \mathrm{L}$ is within a reasonable range of the galaxy, then it indicates that MOND can fit the rotation curve of the galaxy well. So far, many studies using the above method have shown that MOND can fit the rotation curves of many galaxies very well (for instance, in Ref. [21] the author states that no single model can fit all galaxies, but MOND has successfully fitted the largest number of galaxies). However, some other authors also attempted to fix $M / L$ and take $a_{0}$ as a free parameter to fit the observational data of galaxies, regardless of $a_{0}$ being regarded as a universal constant in MOND. For example, in Ref. [22], Randriamampandry and Carignan studied the mass models of 15 nearby dwarf and spiral galaxies and found that galaxies with higher central surface brightness tend to favour higher values of the constant $a_{0}$, and in Ref. [23], Swaters et al. studied a sample of 27 dwarf and low surface brightness galaxies and found that lower surface brightness galaxies tend to have lower $a_{0}$. The two conclusions just coincide with the conclusion we obtained in Sect. 2, i.e., the larger the mass of a galaxy, the higher the value of $A_{0}$ (remember that $A_{0}$ and $a_{0}$ in the equation have the same meaning).

From the analysis of a large amount of literature, there has always been controversy over whether $a_{0}$ is a universal constant in MOND. On the one hand, some literature show that, fixing $a_{0}$ but allowing $M / L$ to be a free parameter to fit the rotation curve of galaxies can indeed fit some observational data very well. On the other hand, when different galaxy samples are used to analyze whether the $a_{0}$ credible intervals for individual galaxies is compatible among these samples, some literature concluded that $a_{0}$ is not a universal constant (for instance, see Ref. [24]), and some other literature take a more aggressive approach, that is, taking $a_{0}$ as a free parameter to fit the observational data, and it is found that $a_{0}$ is related to the surface brightness of galaxies (for instance, the above mentioned Ref. [22] and Ref. [23]). The main reason for such controversy is that, on the one hand, MOND is not a fundamental theory but more of an empirical model, on the other hand, the exclusion of dark matter (if it exists) in the MOND model limits its ability to fit the observational data.

In this paper, we provide another new perspective, that is, the derived $a_{0}$ is not a constant but vary with the mass of galaxy, and the existence of dark matter is also permitted in the model (which can be used to adjust the value of $a_{0}$, since that the value of $a_{0}$ is related to the mass of galaxy or the gravitational potential). Thus, a model that allows dark matter to exist and $a_{0}$ is variable emerges. In the next section, we will see the advantages of this model over MOND in some aspects.

In addition, we know that the MOND model is also under development, such as the EMOND (i.e., Extend MOND), which is first proposed by Zhao and Famaey [20]. This model assumes that $a_{0}$ is related to the gravitational potential, and the stronger the gravitational potential, the larger the value of $a_{0}$, thereby compensating for the drawback that MOND cannot explain the galaxy clusters. The ideological of this model is consistent with the conclusion we have drawn above, although there are differences in mathematical details. We intend to discuss them in the next paper.

\section{Models comparison}
To compare observation data with $\Lambda \mathrm{CDM}$, MOND and Eq. (\ref{eq08}), the Newtonian gravitational acceleration due to the ordinary matter (i.e., not including dark matter) must be separated from the total acceleration. Here we assume the acceleration due to baryons comes from the bulge and the disk (with no dark halo contribution), that is
\begin{equation}\label{eq10}
 a_{B}(r)=\frac{v_{b}^{2}(r)+v_{d}^{2}(r)}{r}
\end{equation}
Where $r$ is the distance from the center of the galaxy, $v_{b}$ and $v_{d}$ are the rotational velocity due to the bulge and disk respectively.

In $\Lambda \mathrm{CDM}$, in addition to the ordinary matter, dark matter is considered, and the function of dark halo is usually expressed as [25]
\begin{equation}\label{eq11}
\rho_{h}(r)=\frac{\rho_{0}}{\frac{r}{h}\left(1+\frac{r^{2}}{h^{2}}\right)} 
\end{equation}
Where $\rho_{0}$ and $h$ are the scale density and scale radius of the dark halo respectively. Then the mass of dark halo can be expressed as
\begin{equation}\label{eq12}
M_{h}(r)=4 \pi \rho_{0}\left[\ln \left(1+\frac{r}{h}\right)-\frac{r}{r+h}\right]
\end{equation}
 
Then it leads to
\begin{equation}\label{eq13}
v_{h}^{2}=\frac{G M_{h}(r)}{r}
\end{equation}

The total acceleration is then
\begin{equation}\label{eq14}
g_{t}=\frac{v_{b}^{2}+v_{d}^{2}+v_{h}^{2}}{r}
\end{equation}

In MOND, there is no dark matter, and the relationship between gravitational acceleration and centripetal acceleration is expressed as Eq. (\ref{eq06}). Since the interpolation function is empirical, here we take the "simple" function [26]
\begin{equation}\label{eq15}
\mu(x)=\frac{x}{1+x}
\end{equation}

\begin{figure}[!ht]
\includegraphics[width=\columnwidth]{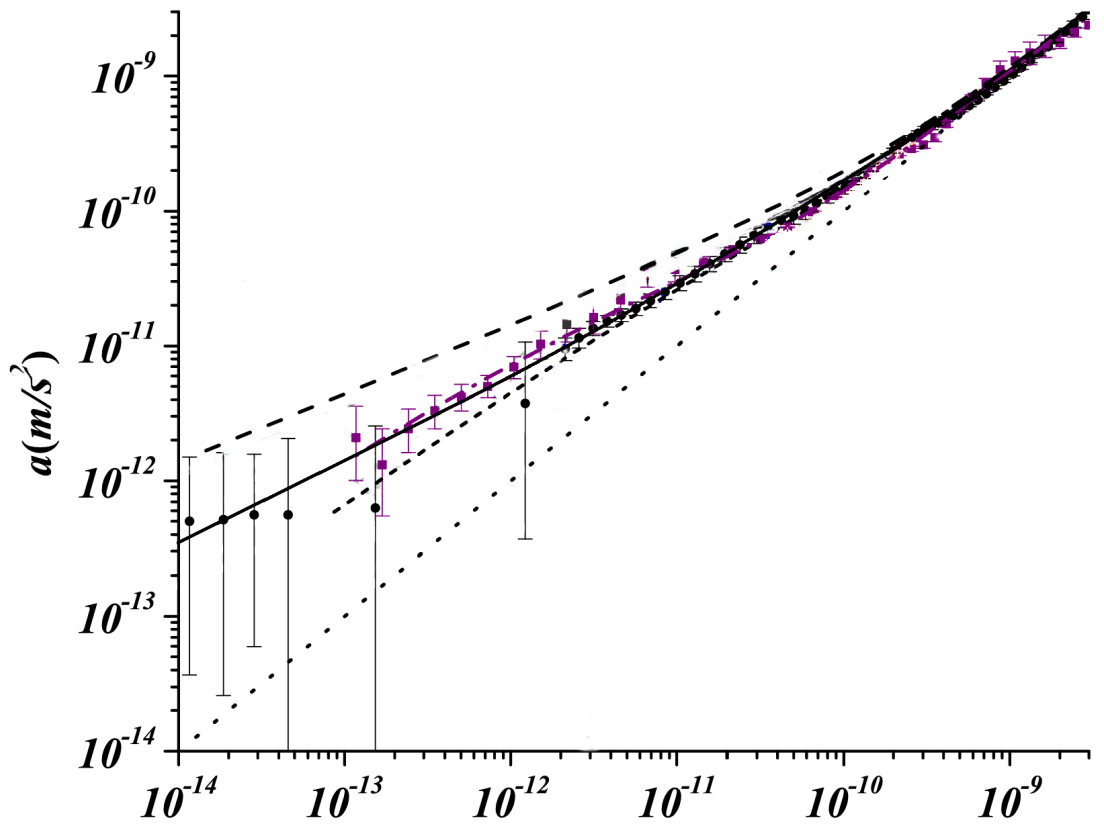}
\caption{The centripetal acceleration vs. the gravitational acceleration due to baryons for data and models. The black circles with error bars represent Milky Way data. The purple squares with error bars represent M31. The short-dashed line is the ACDM fitting curve of Milky Way, the dash-dot line is the $\boldsymbol{\Lambda} \mathbf{C D M}$ fitting curve of M31. The dotted line is the reference line for $\boldsymbol{a}=\boldsymbol{a}_{\boldsymbol{B}}$. The dashed line is predicted by MOND with $\boldsymbol{a}_{\mathbf{0}}=1.2 \times 10^{-10} \boldsymbol{m} / \boldsymbol{s}^{\mathbf{2}}$ and the solid line is predicted by Eq. (\ref{eq08}) with $A_{0}=1.18 \times 10^{-11} \mathrm{~m} / \mathbf{s}^{2}$.}\label{fig02}
\end{figure}

Then we can plot the centripetal acceleration vs. the gravitational acceleration due to baryons for the observation data in Figure \ref{fig02} [27], and the fitting curves corresponding to various models are also plotted.

It should be note that the prediction in Figure \ref{fig02} assumes that the galaxies being observed are isolated. Figure \ref{fig02} shows that for smaller radii, i.e. large accelerations, MOND and Eq. (\ref{eq08}) do not significant deviate from expected ACDM model galaxies, but that changes for lower accelerations. In addition, it indicates that the centripetal accelerations in M31 and Milky Way are smaller than the MOND prediction at lower $a_{B}$, but in line with the prediction of Eq. (\ref{eq08}). This is mainly because there are no free parameters in the prediction of radial acceleration relation (RAR) of galaxies in MOND, while in Eq. (\ref{eq08}), $A_{0}$ is a parameter that can vary with the mass of the galaxy, so we can adjust $A_{0}$ to fit the observation data in Figure \ref{fig02}.

One may wonder whether dark matter was taken into account in Eq. (\ref{eq08}) when fitting the observation data in Figure \ref{fig02}. The answer is that whether you consider it or not makes no mathematical difference to the fitting results. When dark matter exists, we assume that dark matter exists near the center of the disk galaxy (similar to MOND, our model does not need the dark matter halo to be carefully adjusted to meet the galaxy's asymptotically flat rotational velocity curve), which means that the coordinate $a_{\mathrm{B}}$ in Figure \ref{fig02} should be replaced by $g_{t}$. Since dark matter exists near the center of the galaxy, and the radial position corresponding to the maximum value of $a_{\mathrm{B}}$ in Figure \ref{fig02} is far away from the center of the galaxy, it is easy to prove that the $g_{t}$-axis can be obtained by adding a same value (which is determined by the mass of dark matter) to all values of the $a_{\mathrm{B}}$-axis, or in other words, the whole shift of the $a_{\mathrm{B}}$-axis to the right side can yield the $g_{t}$-axis. In Eq. (\ref{eq08}), the left side is just related to $g_{t}$, and the right side is just related to $a$. In addition, from Tab. 1 it has been pointed out that the function form of $\mu^{\prime}$ has nothing to do with the central mass of the galaxy (that is, $\mu^{\prime}$ is independent of $a_{\mathrm{B}}$ and $g_{t}$ ), so it can be easily proved that whether there exists dark matter near the center of the galaxy does not affect the fitting result of Eq. (\ref{eq08}) mathematically. But physically, dark matter may be needed as its mass can help to adjust the value of $A_{0}$.

\section{Discussion on $\boldsymbol{\lambda}$}
Obviously, the value of $\lambda$ assumed above is crucial to the evolution of old galaxies (but not so important for young ones). Then, an important question is, what is the value of $\lambda$ ? Or, as a second choice, what order of magnitude might the value of $\lambda$ be. As there is currently no clear physical basis for $\lambda$, so at present the specific value of $\lambda$ is difficult to obtain. But we can roughly obtain the order of magnitude of $\lambda$ from the rotation curve of the galaxy.
\begin{figure}[!ht]
\includegraphics[width=\columnwidth]{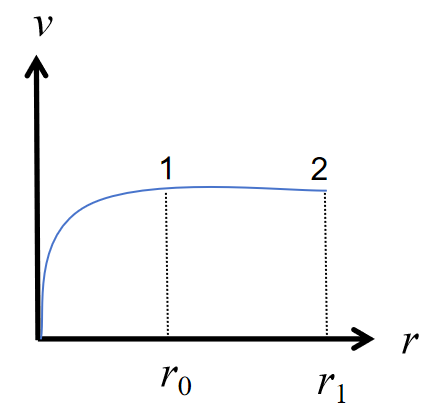}
\caption{The asymptotically flat rotational curves presented by many galaxies.}\label{fig03}
\end{figure}

From Eq. (\ref{eq04}) it has $v r=v_{0} r_{0} \mathrm{e}^{\lambda t}$, then when $v=v_{0}$, there is $r=r_{0} \mathrm{e}^{\lambda t}$, which indicates that the radial coordinates of a rotating object will continuously increase over time (but its circumferential rotational velocity remains the same). And we know that the rotation curves of many galaxies are asymptotically flat, as shown in Figure \ref{fig03}, and from the viewpoint of Eq. (\ref{eq02}) or Eq. (\ref{eq05}), in Figure \ref{fig03}, when point 2 and point 1 are on the flat curve, point 2 is the evolution of point 1 over time.

Therefore, we can calculate $\lambda$ based on the radial coordinate ratio of point 1 to point 2 (i.e., $r_{1} / r_{0}$ ) as well as the time required for point 1 to evolve to point 2 .

In Ref. [28], many rotational velocity curves of galaxies are listed (all of which show the characteristic of asymptotically flat), from these rotational velocity curves we can find that $1<r_{1} / r_{0} \leq 4$, however, the time for evolution from point 1 to point 2 is hard to obtain, which is not only related to the mass of the galaxy, but also to the initial orbit of the object and the value of $\lambda$. Here, we take the timescale of the universe's evolution to roughly estimate the order of magnitude of $\lambda$, that is, we assume $t=10 \mathrm{Gyr}$, then we can obtain that the order of magnitude of $\lambda$ is about $10^{-18} / s$, which coincidentally is the same as the order of magnitude of the present Hubble constant. It reminds us that, with such a small value of $\lambda$, the radial expansion of the galaxy itself caused by it is also extremely small, which brings great difficulties to the detection.

Another question, one may wonder whether $\lambda$ must be a constant. As has been stated that, there is currently no clear physical basis for $\lambda$ (Perhaps it originated from an unknown but deeper theory), so we don't know the answer at present, however, if the rotational velocity curve of the galaxy does eventually tend to be "absolutely" flat, then $\lambda$ must be a constant.

\section{Summary}
This paper, with no clear physical basis, just follows the idea in Ref. [16] with curious, that is, introducing a viscosity coefficient into the $\Lambda \mathrm{CDM}$ model to describe the expansion of universe, also introduces such a similar positive viscosity coefficient into the rotational motion describing the disk galaxies, and then studies what will happen. We were surprised to find that it yields all the formulas assumed in MOND, including a concrete interpolation function between the centripetal acceleration and the Newtonian acceleration, which however is empirical in MOND.

But we also obtain something different from MOND. First, $a_{0}$ is considered to be a constant in MOND, but we obtained that $A_{0}$ (here it just to distinguish from $a_{0}$ in MOND, and $A_{0}$ and $a_{0}$ in the equation have the same meaning) is not a constant but is related to the mass of the galaxy, and the more massive the galaxy, the greater the value of $A_{0}$ if the value of $\lambda$ is a constant. For this theoretical prediction, we find that several previous studies have already drawn consistent inferences based on observational data. This characteristic can help it to fit the RAR data better than MOND. Second, the paper predicts that the rotational galaxies will expand radially over time at an extreme small expansion rate, just like the expansion of universe, so it predicts a "dynamical" galaxy rather than a "static" galaxy predicted by $\Lambda \mathrm{CDM}$ and MOND. Currently, observational evidence for this theoretical prediction remains extremely limited. Only one recent study has reported tentative signatures of galactic expansion based on the Gaia DR3 dataset [29].

In many literature, such as Refs. [30-32], it is found that MOND can fit the observation data of rotational galaxies better than $\Lambda \mathrm{CDM}$, which is thought to be mainly attributed to the two extreme conditions (i.e., Newtonian and deep-MOND limits) in MOND, and has no or little relationship with the concrete form of the interpolation function in MOND. For these observation data, it is easy to know that the model in this paper can also fit them well as the two extreme conditions can also be naturally derived from our model rather than be assumed in MOND. In MOND, in order to fit these observational data, the mass-to-light ratio of the galaxy, which is the only free parameter in MOND [33], have to be adjusted, and therefore the existence of dark matter must be excluded, because the presence of dark matter can affect the mass-to-light ratio. However, the model in this paper cannot rule out the existence of dark matter, one reason is that the mass of dark matter can help to adjust the value of $A_{0}$ to better fit the RAR curves, and another reason is that the idea of this paper comes from Ref. [16], which supports the existence of dark matter. However, unlike $\Lambda \mathrm{CDM}$, in our model, the asymptotically flat rotational velocity curve of the disk galaxy does not require the careful adjustment of dark matter (if it exists) to be achieved, which seems unnatural. It can be also achieved naturally under the viscous dynamics of the galaxy itself. So, in fact, we tend to consider that the idea of this paper to be a further optimization of $\Lambda \mathrm{CDM}$ rather than a replacement.


\section*{Data Availability Statement}
No Data associated in the manuscript.



\vspace{-1em}






\end{document}